\documentclass[aps, prl, twocolumn, superscriptaddress, amsfonts, amssymb, amsmath]{revtex4}

\usepackage{graphicx}

\begin{document}

\title{Predicting the rise of right-wing populism in response to unbalanced immigration}

\author{Boris~Podobnik}
\affiliation{Boston University, Boston, Massachusetts 02215, USA}
\affiliation{University of Rijeka, 51\,000 Rijeka, Croatia}
\affiliation{Zagreb School of Economics and Management, 10\,000 Zagreb, Croatia}
\affiliation{Luxembourg School of Business, Luxembourg}
\author{Marko~Jusup}
\affiliation{Center of Mathematics for Social Creativity, Hokkaido University, Sapporo 060-0808, Japan}
\author{H.~Eugene~Stanley}
\affiliation{Boston University, Boston, Massachusetts 02215, USA}

\maketitle

\textbf{Among the central tenets of globalization is free migration of labor. Although much has been written about its benefits, little is known about the limitations of globalization, including how immigration affects the anti-globalist sentiment. Analyzing polls data, we find that over the last three years in a group of EU countries affected by the recent migrant crisis, the percentage of right-wing (RW) populist voters in a given country depends on the prevalence of immigrants in this country's population and the total immigration inflow into the entire EU. The latter is likely due to the EU resembling a supranational state, where the lack of inner borders causes that "somebody else's problem" easily turns into "my problem". We further find that the increase in the percentage of RW voters substantially surpasses the immigration inflow, implying that if this process continues, RW populism may democratically prevail and eventually lead to a demise of globalization. We present evidence for tipping points in relation to the rise of RW populism. Finally, we model these empirical findings using a complex network framework wherein the success of globalization largely rests on the balance between immigration and immigrant integration.}

\section{Introduction}~An important aim of globalization is ensuring that, alongside capital, labor moves freely across national borders~\cite{Grubel, Docquier, Hufbauer, Dani16}. In the EU, therefore, a country lacking labor force should welcome laborers from another EU country with excess labor. This economic argument is somewhat negligent of how such a welcoming policy may affect public opinion, expressed in democratic societies via popular vote. Especially volatile situations may arise if either the native majority or the migrant minority sense that their national or religious identity is being threatened. In this context, an unprecedented inflow of immigrants into the EU during the recent migrant crisis seems to offer a fresh insight into the relationship between popular vote and immigration and, by extension, into the factors that directly affect the success of further globalization.

Although a large body of literature is dedicated to the analysis of how migration affects the global economy~\cite{Grubel, Docquier, Hufbauer, Chiswick, Borjas, Borjas1, Borjas2, Swank, Gibson} and right-wing (RW) populism~\cite{Betz, Knight, Fysh, Voerman, Chapin, Smith, Bonikowski}, much less is known about the limitations of globalization~\cite{Dani00, Dani16}, especially how large-scale migrations sway popular vote and what the economic consequences may be. Borjas reported slow integration of immigrants into the US, taking four generations to catch up with the earnings of natives instead of one or two as commonly believed~\cite{Borjas3}. In a slightly broader context, particularly interesting is the relation of immigration to globalization. Rodrik's argument~\cite{Dani00, Dani16} states that globalization, democracy, and national sovereignty are mutually irreconcilable, leading to a conclusion that democracy is compatible with national sovereignty only if globalization is restricted.

The recent migrant crisis in the EU was motivated by political turmoil and armed confrontations rather than globalization itself, yet as an unintended consequence some of the central tenets of globalization have been put to test. In particular, the increase of immigrants in the total population was paralleled with a surge in voters who support RW populist parties. The growing number of RW voters across the EU suggests that tolerance towards immigrants is conditional~\cite{BP16} because parts of the general public with little regards for RW populism at first, demonstrably turned into RW populist supporters. Understanding this "change of heart" is of utmost importance for the success of globalization. If the immigration inflow exceeds the speed of integration for a prolonged period of time~\cite{BP16}, the RW populist movement may democratically prevail and eventually lead to a demise of globalization. Even if one is not concerned with globalization, humanity needs to better prepare for the potential massive displacements of the global population expected due to global climate change.
  
Herein we find that during the last three years, in the EU countries hit by the recent migrant crisis, the speed of immigrant inflow substantially exceeds the speed of their integration. We further find a significant relationship between the percentage of RW voters and the fraction of immigrants, on the one hand, as well as the total immigration inflow into the entire EU, on the other hand. By contrast, injuries and casualties in violent incidents and the integration rate proved to be poor predictors of the percentage of RW voters. Seeing globalization as a tolerant mode of democracy, wherein cooperation between nations supplements national interest, an important question arises naturally: Can we predict under what circumstances the RW populist movement receives enough support to overthrow this tolerant mode of democracy, potentially leading to drastic political and economic upheaval? Moreover, because globalization makes many countries experience similar problems, should we expect a cascade (i.e., a domino) effect, whereby the rise of a populist movement in one country triggers similar movements in other countries?
  
\section{Results}~RW populism is often characterized by intolerance which, in turn, is among the most widespread social phenomena, commonly responsible for conflicts and segregation \cite{Schelling, Axelrod97, McPherson, Antal, Marvel, Moreno, Parravano}. Together with the associated phenomenon of radicalization, intolerance is the main cause of violence and terrorism \cite{Lim, Dancygier1, Dancygier2, Galam, Sampson}. Here, we use the fraction of RW populist voters in a given country as a proxy for RW populism. For each of the countries most affected by the recent migrant crisis, we estimate the percentage of immigrants from September 2013 to June 2016, starting from the official value in 2013 and complementing it with the number of monthly recorded visa applicants \cite{inflow}. For the same time period, we also collect the available election polls and election results \cite{DATApolls}.

\begin{figure}[t]
\centering \includegraphics[scale=0.80]{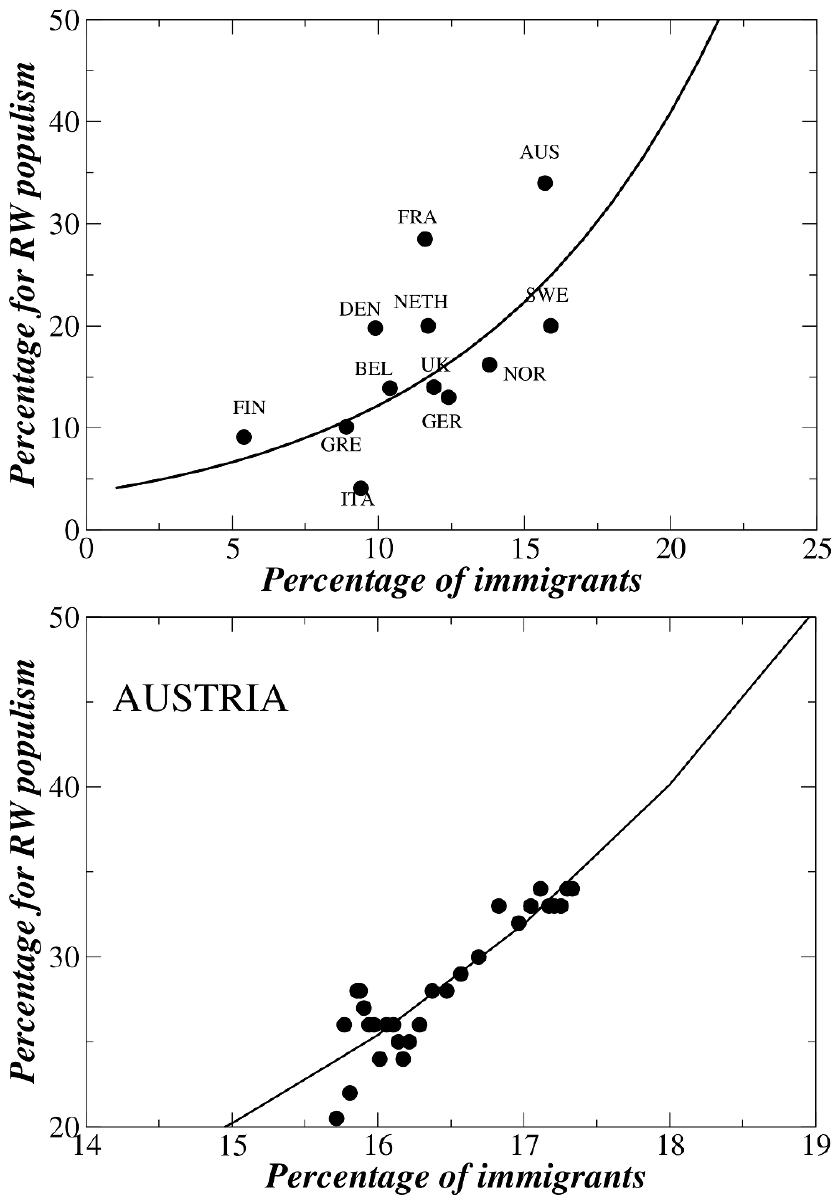}
\caption{\textbf{Immigration affects the support for right-wing populism I.} (a) Among the EU countries involved in the recent migrant crisis, support for RW populism is generally higher in those countries that accepted a larger number of immigrants relative to the country's population size. Shown is June 2016. Seeing democracy as the majority rule principle, we presume that RW populism becomes a dominant political option when the percentage of RW voters exceeds 50\%. Judging based on a cumulative exponential function that fits the data reasonably well, RW populism in the examined EU countries may take over if the percentage of immigrants in the total population approaches 22\%. (b) Similar as in the other EU countries, Austrian data reveal that the increase in the percentage  of immigrants is accompanied with an increase in the percentage of RW populist voters. Here too a cumulative exponential function fits the data well. This function predicts the rise of RW populism in Austria when the percentage of immigrants is slightly below the 20\% mark.}
\label{f1}
\end{figure}

The situation in the EU in June 2016 is summarized in Fig.~\ref{f1}(a) that reveals a rising trend in the percentage of RW populist supporters in response to the increasing percentage of immigrants in the general population. From the cumulative exponential function that fits the data well, we find that as the percentage of immigrants approaches approximately 22\%, the percentage of RW populist voters exceeds 50\%, a threshold needed in democratic societies for any party, and thus RW parties too, to take over the government. Considerable scatter in the data furthermore suggests that there may be visible differences in (in)tolerance between the countries---the larger the percentage of voters in favor of RW populism compared to the percentage of immigrants, the smaller the tolerance. Taking Austria as an example, in Fig.~\ref{f1}(b) we find that the 50\% threshold would be reached even if the percentage of immigrants was below 20\%.

\begin{figure}[t]
\centering \includegraphics[scale=0.80]{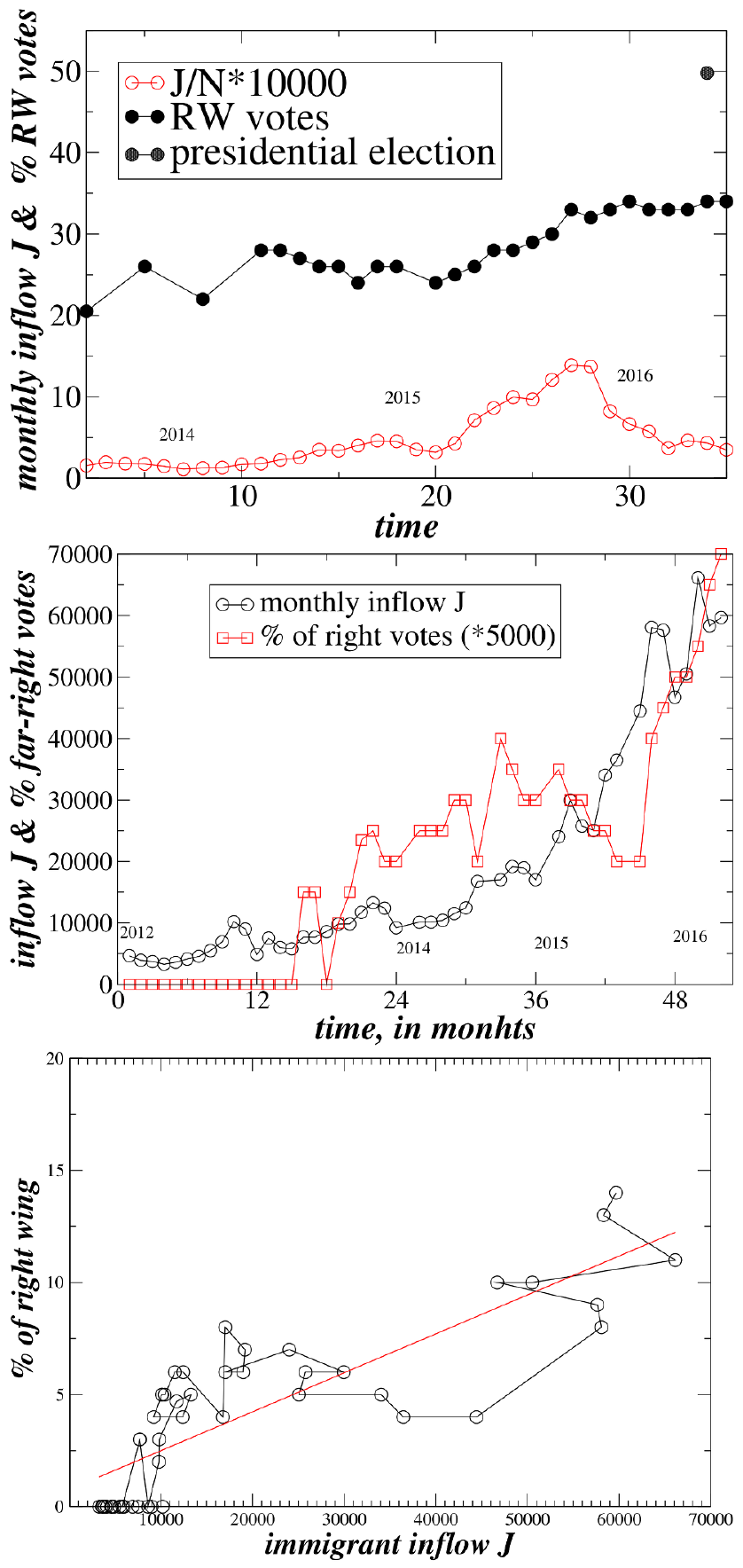}
\caption{\textbf{Immigration affects the support for right-wing populism II.} (a) An unprecedented inflow of immigrants into Austria coincided with a steady increase in the fraction of RW populist voters. 
A solitary black dot represents the results of Austrian presidential election in May 2016 in which an RW populist candidate secured almost 50\% of votes. This election shows that even after the record immigrant inflow at the end of 2015 had subsided, a decreasing trend in the number of immigrants that enter Austria did not automatically translate into lower support for the RW populist political option, i.e., RW populism seems to be more than just a craze. (b) In Germany, the increasing inflow of immigrants (monthly data~\cite{inflow}) rather clearly coincided with the increasing support for an RW populist party. (c) A significant regression emerges when the German case is presented as a scatter plot between the inflow of immigrants and the percentage of far-right voters.}
\label{f2}
\end{figure}

The fraction of immigrants in the general population is not the only factor affecting the sentiment of voters. In Fig.~\ref{f2}, using the data for Austria and Germany over the past three years (2013-2016), we demonstrate that the percentage of RW populist supporters also depends on the inflow of immigrants into Europe. Illustrative is the Austrian example, where in 2013 parliamentary election the far-right party won 20.5\% of the popular vote, roughly reflecting the sentiment predicted from the percentage of immigrants living in Austria at the time. However, due to a high inflow of immigrants that in the second half of 2015 reached unprecedented proportions~\cite{inflow}, the local Vienna election saw the percentage of RW voter suddenly jump to 33\%. This sudden change in popular vote is reminiscent of phase transitions (i.e., tipping or critical points)---well documented in social sciences~\cite{PRL, Mas}---whereby the closer a country to a tipping point, the more abruptly voters turn their back to moderate parties and start voting for more extreme alternatives. A qualitatively similar phenomenon is seen in the case of Germany in Fig.~\ref{f2}(b)-(c).  

\begin{figure}[t]
\centering \includegraphics[scale=0.80]{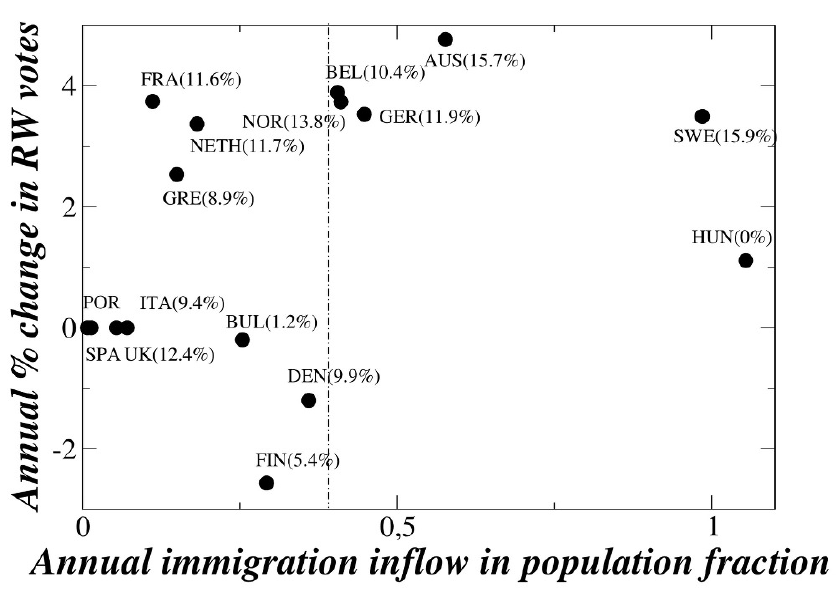}
\caption{\textbf{Immigrant inflows and the popularity of right-wing populist movements---a non-linear threshold.} Shown is the annualized immigrant inflow into a given country (horizontal axis) as a percentage of that country's population, as well as the corresponding percentage change in RW populist votes (vertical axis). In parentheses are the fractions of immigrants in the total population of the corresponding country. For a group of countries in which the annualized increase in the percentage of RW voters exceeded 2\%, this increase is virtually independent on the inflow of immigrants. Such a result may reflect the EU's political organization, i.e., the lack of internal borders whereby if one country decides to accept immigrants, the decision may have repercussions for all the other member states. We also observe a threshold indicated by a dashed line at which the immigrant inflow into a given country is sufficiently high to invariably provoke an increase in the percentage of RW populist voters. In the model construction, this threshold suggests $\alpha=0.004$ on a annual basis.}
\label{f3}
\end{figure}
 
In an attempt to probe deeper into the internal dynamics of RW populism in the EU as a function of the inflow of immigrants, next we analyze how the immigration rate affects the rise in RW populist voters. Surprisingly, for a group of countries in which the annualized increase in the percentage of RW voters exceeded 2\%, Fig.~\ref{f3} shows that this increase is virtually independent on the inflow of immigrants. Why would countries with a relatively high and a relatively low inflow of immigrants exhibit about the same increase in the percentage of RW voters? This result may be a consequence of the EU's political organization. Because the EU functions practically as a supranational state with no internal borders, if one country decides to accept immigrants, this decision may have repercussions for all the other member states. The increase in the percentage of RW populist voters may therefore more systematically depend on the total inflow of immigrants into the entire EU, expressed here as a percentage of the total EU population, than the inflow in any individual country.

Some, albeit anecdotal, evidence to the effect that the decision of one country may affect the situation in another is seen in the case of Sweden and Norway. The former country was among those that were hit the hardest by the recent migrant crisis, yet the latter country saw practically the same annualized increase in the percentage of RW voters. Another interesting pair in this context is Germany and Poland. Again it was the former country that experienced a high inflow of immigrants, yet it is in Poland that 53\% of the population thinks that their government should refuse asylum seekers from the Middle East and North Africa (and only 33\% thinking Poland should do the opposite). The Polish example may contain another important lesson. Namely, this country seems to have already transitioned from the tolerant mode of democracy associated with globalization to a mode dominated by RW populism. If so, the implication is that the fraction of immigrants at which the Polish population is pushed beyond the tipping point is much lower than in western EU countries. Poland---and similarly Hungary, both of which share decades of socialist experience---is among the toughest opponents of immigration into the EU, strongly debating against the quotas that the EU imposed with a goal to more evenly spread the shock of recent migrant crisis. 
 
Because Fig.~\ref{f3} implies that the interplay of factors influencing the popularity of RW populism is more complex than a simple bivariate regression could reveal, we turn to econometric analysis and multivariate regression. Based on the results of simple regression in Fig.~\ref{f1}, we take that the fraction of RW voters (response variable, $RW_t$) in a given country is primarily determined by the fraction of immigrants ($IM^L_t$) in this country. To account for the assumption that the percentage of RW voters is affected by the overall inflow of immigrants into the EU ($IM^{EU}_t$), this variable is also included in the model. Finally, to control for the possibility that violent incidents involving immigrants may sway the popular vote, we include into the model the total number of injuries ($I$) and casualties ($D$) recorded in such incidents across the EU~\cite{wiki}. Thus, the regression model becomes
\begin{equation} 
RW_t = \beta_0 + \beta_{IM}^L IM^L_t + \beta_{IM}^{EU} IM^{EU}_t + \beta^D_{ter} D_t + \beta^I_{ter} I_t + e_t, 
\label{e1}
\end{equation} 
where $e_t$ is random error. In three countries that were hit the hardest by the recent migrant crisis (Germany, Austria, and Sweden), we find a significant relationship between the fraction of RW voters and the fraction of immigrants (Table~\ref{t1}). In Germany and Sweden, furthermore, the support for RW populism is positively related to the immigration inflow into the EU. Somewhat surprisingly, there is no significant dependence of the response variable on the number of injuries and casualties in violent incidents.

\begin{table}[t]
\begin{center}
\tabcolsep=0.11cm
\begin{tabular}{ccccccc}
\textit{state} & ${\beta_0}$ & ${\beta^{L}_{IM}}$ & ${\beta^{EU}_{IM}}$ & ${\beta_{ter}^{D}}$ & ${\beta_{ter}^{I}}$ & ${R^2}$ \\
\hline
\textit{Ger}   & -1.17$^*$   & 10.29$^*$          & -4.3e-07$^*$        & 1.4e-04             & 1.563e-05           & 0.852   \\
$ $            & (-11.82)    & (12.31)            & (-5.36)             & (0.67)              & (0.19)              &         \\
\hline
\textit{Aus}   & -1.47$^*$   & 10.11$^*$          & 7.9e-07             & -3.1e-04            & 2.1e-05             & 0.452   \\
$ $            & (-2.11)     & (2.24)             & (1.23)              & (-0.20)             & (0.03)              &         \\
\hline
\textit{Swe}   & -0.34$^*$   & 2.60$^*$           & 5.8e-07$^*$         & -1.9e-04            & 2.1e-05             & 0.751   \\
$ $            & (-2.88)     & (3.53)             & (3.37)              & (-0.47)             & (0.13)              &         \\
\hline
\hline	
\end{tabular}
\end{center}
\caption{Multiple regression as defined in Eq.~(\ref{e1}) using the data on the three EU countries most gravely affected by the recent migrant crisis. Star denotes the statistically significant regression coefficients at the 5\% significance level. Parentheses hold the value of t-statistic.\label{t1}}
\end{table}

\begin{table}[t]
\begin{center}
\tabcolsep=0.11cm
\begin{tabular}{ccccc}
$ $               & {Coeff.} & {Std. Err.} & {t Stat.} & $P>t$ \\
\hline
$\beta_{IM}^{L}$  & 3.48     & 0.486       & 7.17      & 0.000 \\
\hline
$\beta_{IM}^{EU}$ & 126.4    & 41.6        & 3.04      & 0.002 \\
\hline
$\beta_{ter}^D$   & 8.3e-05  & 7.7e-05     & 1.07      & 0.284 \\
\hline
$\beta_{ter}^I$   & -1.2e-04 & 1.9e-04     & -0.64     & 0.525 \\
\hline
\textit{MIPEX}    & 0.120    & 0.244       & 0.49      & 0.622 \\
\hline
$\beta_0$         & -0.353   & 0.109       & -3.24     & 0.001 \\
\hline
\hline	
\end{tabular}
\end{center}
\caption{Pooled time series cross-section analysis (TSCS) with random-effects GLS regression as defined in Eq.~(\ref{e2}). Test statistics: 
Wald $\chi^2(5)=139.39$ and $Prob > \chi^2=0.000$.\label{t2}}
\end{table}

\begin{table}[t]
\begin{center}
\tabcolsep=0.11cm
\begin{tabular}{ccccc}
$ $               & {Coeff.} & {Std. Err.} & {t Stat.} & $P>t$ \\
\hline
$\beta_{IM}^{L}$  & 3.99     & 0.557       & 7.18      & 0.000 \\
\hline
$\beta_{IM}^{EU}$ & 163.2    & 42.1        & 3.88      & 0.000 \\
\hline
$\beta_0$         & -0.432   & 0.089       & -4.84     & 0.000 \\
\hline
\hline	
\end{tabular}
\end{center}
\caption{Pooled time series cross-section analysis (TSCS) with random-effects GLS regression. Test statistics: Wald $\chi^2(5)=169.9$ and $Prob > \chi^2=0.000$.\label{t3}}
\end{table}

We extend the above econometric analysis with a pooled time-series cross-section (TSCS) method that combines the cross-sectional data on multiple countries. Here, the number of countries is $N=7$: Germany, Austria, the Netherlands, Sweden, France, Norway, and Denmark. Because for each country there are $T_i$ observations along the temporal dimension, the whole dataset has $\sum_{i=1}^N T_i=142$ 
observations. Compared to the model in Eq.~(\ref{e1}), the TSCS model has an extra variable, $M_{it}=(1 - MIPEX/100)$, where MIPEX is the Migrant Integration Policy Index (MIPEX)~\cite{mipex}, a proxy for the integration rate---a higher MIPEX implies better integration---and an extra index $i=1, 2,..., N$ that refers to a cross-sectional unit:
\begin{equation} 
RW_{it} = \beta_0 + \beta_{IM}^L IM^L_{it} + \beta_{IM}^{EU} IM^{EU}_{it} + \beta^D_{ter} D_{it} + \beta^I_{ter} I_{it} + M_{it} +  e_{it}.
\label{e2}
\end{equation} 
The results of the TSCS regression model emphasize the fraction of immigrants in the general population and the inflow of immigrants entering the entire EU as the significant predictor variables (Table~\ref{t2}). The results of the TSCS regression model without MIPEX and violent incidents are shown in Table~\ref{t3}. Interestingly, even if we neglect the total inflow of immigrants into the EU, coefficient values in Table~\ref{t3} suggest that between 23\% and 24\% of immigrants in the total population of a country are sufficient to cause larger than 50\% support for RW populist parties, which is similar to the result obtained from Fig.~\ref{f1}.

Based on the analyzed data it is unclear whether the rise of RW populism is just a transient phenomenon or if a change in political orientation is accompanied with longer term memory. A glimpse into the persistence of voters' memory is offered by the sequence of events in Austria following the above-mentioned local Vienna election held in October 2015 wherein the RW populist party won 33\% of the popular vote. Namely, during the presidential election a few months later, just as the migrant crisis was at its peak, the RW populist movement received another boost and its candidate secured almost 50\% of votes, narrowly losing to a leftist rival. However, these results were contested and the re-vote is supposed to take place in early December 2016. Despite the migrant crisis having subsided since, polls have the RW populist party holding steady at around 35\% and its candidate at almost 51\%, thus indicating that the rise of RW populism is indeed more than just a craze. This indication seems to have some support in the literature as well. A substantial increase in the number of refugees and illegal immigrants in European countries during the 1980s provoked a similar wave of radical RW populism as described herein~\cite{Betz93}. In the wake of these events, already in the early 1990s, there were still between 11 and 14\% of the Europeans who categorized other nationalities, races, or religions as unsettling~\cite{Betz93}.

\section{Model}~Human interactions are for many purposes heterogeneous and prone to abrupt non-linear responses. Precisely such a response is seen in the rise of the RW populist party and its candidate in the Austrian elections. Accordingly, linear approaches as exemplified by regression Eq.~(\ref{e1}) are bound to get us only so far. To offer a mechanistic perspective on the rise of RW populism and account for the existence of tipping points in social dynamics, we adopt a complex network approach in the spirit of Refs.~\cite{BP16, NP14, BP15, Lee15}. The benefit of relying on complex networks, due to their ability to emulate the stated heterogeneity in human interactions, goes beyond just capturing the dynamics in the vicinity of tipping points. Heterogeneity seems important when considering immigration and integration issues because integrating immigrants who live in ghettos (or hubs in a network-theoretic jargon) may be substantially more difficult than if immigrants mixed uniformly with the native population.

In accordance with our interest in the increase of the immigrant population relative to the native one, without any loss of generality, we begin the model construction by setting a constant number of native (hereafter also insider) agents. These agents are arranged in an Erd\H{o}s-R\'{e}nyi random network of business and personal contacts. Immigrants (hereafter often outsiders) populate the network subsequently. Every insider observes the fraction of outsiders in their neighborhood and based on this fraction decides whether to be supportive of globalization or to turn to populism. Because insider agents refer to their neighborhood for information, the interaction is local, but also essential for generating and understanding tipping points~\cite{Watts}. There are, however, other non-local types of interactions, as well as the possibility that information is misperceived or misinterpreted. In the following, we formalize these concepts with three model assumptions. 

\paragraph{Assumption (i): media and economic influences.} At each time $t$, representing a period of one month, every insider agent is influenced by media at relative rate $p$ (i.e., a probability per unit of time) and stays influenced for period $\tau$. We assume that this influence turns insiders into RW populist supporters. Again there is no loss of generality because, although media affect insiders both ways, we are interested only in the net rate (negative $p$ would indicate the opposite effect). The effect of media is global and reflects the possibility that just hearing about immigration may turn some insiders into RW populist supporters whatever the true local situation. More generally, rate $p$ could also reflect economic factors (e.g., unemployment). For a randomly chosen insider agent, we readily calculate the probability that this agent is influenced by media using equation~\cite{NP14} $p*=1-\exp(-p\tau)$.
 
\paragraph{Assumption (ii): local influence of outsiders.} In Greek local elections in Nov 2010, a far-right party Golden Dawn got 5.3\% of the vote, yet in some neighborhoods of Athens with large immigrant communities, the party won up to 20\%. This kind of election results suggest that contacts between insiders and outsiders matter. In our network model, a constant number of $N$ insiders is supplemented with $I(0)$ outsiders initially, where the latter number increases at each moment $t$ due to inflow $J_t$. Newly arriving outsiders are placed randomly between existing agents who at the beginning have an average number of connections (i.e., degree) $k$. The total number of outsiders, $I(t)$, is obtained by summing monthly $J$'s according to $I(t)=I(0)+\Sigma_{s=1}^t J_s$. At any moment, the fraction of outsiders equals $f_I(t)=I(t)/(N+I(t))$. To account for the above-mentioned effect due to contacts between insiders and outsiders, we assume that any agent $i$ with $k_i$ total connections turns to RW populism at rate $p'$ if this agent is surrounded by at least $m_i=f_I' k_i$ outsiders~\cite{Watts, NP14, BP15b}, where $0<f_I'<1$ is a constant model parameter quantifying how tolerant insiders are. This assumption merits a few additional comments.

First, the probability that randomly chosen insider agent $i$ with $k_i$ connections is surrounded by $m_i$ outsiders, and therefore prone to RW populism, equals $p_1(k_i,m_i,f_I)\equiv\Sigma_{j=m_i}^{k_i} f_I^{j}(1-f_I)^{k_i-j}{k_i\choose k_i-j}$. In this formula, $f_I$ is the true current state of the network. The reality may, however, be such that information is biased and insiders perceive more outsiders than there really are. If the bias is $\Delta f_I$, then $p_1(k_i,m_i,f_I+\Delta f_I)>p(k_i,m_i,f_I)$. This increased probability $p(k_i,m_i,f_I+\Delta f_I)$ implies that the tolerance parameter, $f_I'$, must decrease by amount $\Delta f_I'$ estimable using condition $p(k_i,m_i-\Delta f_I' k_i,f_I)=p(k_i,m_i,f_I+\Delta f_I)$. An implicit assumption here is that all insider agents are equally tolerant to immigrants because the tolerance parameter, $f_I'$, is defined as a global network property rather than an individual agent property. An alternative would be to assume a distribution of tolerance in which case $f_I'$ represents the mean~\cite{BP16}.

To accommodate the econometric results that immigration inflows affect the support for RW populism, we extend assumption (ii) by requiring that (A) if the immigration inflow, $J$, is below some threshold $J'$, society is gradually becoming more tolerant. Consequently, if $J<J'$ at a given time moment, then the tolerance parameter, $f_I'$, increases at this moment by amount $\Delta f_I'=\delta>0$. When extension (A) is met, there is a balance  between immigration and immigrant integration---outsiders are successfully integrated---and insiders are capable to acclimate to changes in their society. Judging naively from Fig.~\ref{f2}(b), $J'$ for Germany should be close to 10\,000 people per month. The problem of finding a maximum $J'$ at which insiders acclimate to an increasing presence of outsiders may be interpreted as highly relevant for the success of globalization. Namely, in the context of his trilemma set, Rodrik argued~\cite{Dani00, Dani16} that the notion of democracy is compatible with national sovereignty only if globalization proceeds in a carefully balanced manner. One such balance, achievable in the model if $J<J'$, concerns immigration and integration as inevitable consequences of globalization. Otherwise, further progress is threatened by the rise of RW populism as an intolerant mode of democracy wherein cooperation between nations is, to put it mildly, downgraded on the list of political priorities.

Empirical evidence suggests that in addition to extension (A), we should consider an opposite extension (B). Specifically, as the inflow of outsiders, $J$, crosses some threshold $J''$ (not necessarily equal to $J'$), society gradually becomes less tolerant. A mathematical representation of extension (B) would be that if $J>J''$, then the tolerance parameter, $f_I'$, decreases by amount
\begin{equation}  
\Delta f'_I=-\gamma J. 
\label{e3}
\end{equation}
In view of the econometric models in Eqs.~(\ref{e1}) and (\ref{e2}), we make the decrease of the tolerance parameter proportional to inflow $J$, where $\gamma$ is a proportionality coefficient expressing the sensitivity of insiders to large outsider inflows. Furthermore, motivated by Fig.~\ref{f3}(b), we express threshold $J''$ in terms of the total population, i.e., $J''=\alpha N$, where $\alpha$ is another proportionality coefficient. Fig.~\ref{f3}(b), in fact, hints what an estimate of $\alpha$ would be in the case of the EU countries hit by the recent migrant crisis. The dashed line in this figure delineates the annual inflow below which countries responded to immigration in a mixed fashion, but above which support for RW populism always increased. Because $\alpha=J''/N$ and, based on Fig.~\ref{f3}(b), $12J''/N\approx 0.004$, we obtain
\begin{equation}  
\alpha\approx 0.00033.
\label{e4}
\end{equation}

Extensions (A) and (B) represent two opposite limiting cases, one in which immigration is a relatively slow process and the other in which immigration is a relatively fast process. We were motivated to introduce these two extensions by the empirical evidence, yet some support can be found in other sciences too. Brain science, for example, offers a physiological interpretation stating that political attitudes have a counterpart in the brain structure~\cite{Kanai11, Zamboni09, Oxley08}. If outsiders increase in a manner perceived as controlled by insiders, the prefrontal cortex, a part of human brain responsible for decision making and for moderating social behavior, acclimates to the new circumstances. If, however, insiders start perceiving outsiders as invaders, the prefrontal cortex is overcome by the amygdala which in turn induces a fighting reaction and thus suppresses tolerance. 

The model development so far, via assumptions (i) and (ii), accounts for processes that affect an individual insider's opinion. The rise of RW populism, however, would be most explosive if insiders could affect each other's opinions. Influence by peers is a well-documented phenomenon in human interactions, further catalyzed by the popularity of social media. Accordingly, RW populism has the potential to spread in a highly non-linear way, much like a virulent contagious disease. We include such a non-linear collective spreading mechanism in the third (and last) model assumption.
 
\paragraph{Assumption (iii): mutual insider contagion.} At any given moment $t$, an insider agent $i$ with $K_i$ connections to other insiders, turns to RW populism at rate $p''$ if at time $t-1$ this agent had at least $M_i=K_i/2$ RW populist supporters in the immediate neighborhood. Note that for simplicity, factor 1/2 plays a role analogous to the tolerance parameter in assumption (ii). The essence here is that, due to connections between insider agents, once RW populism emerges somewhere in the network, the whole populist movement can spread like a contagion. Such a collective spreading is essential because an insider agent can become RW populist supporter even if there are no outsiders in the immediate neighborhood, thus potentially explaining why some EU countries with almost no immigration oppose to accept even a small group of immigrants. A similar effect may have been important for the results of the recent presidential election in the US wherein the winning candidate was more often than not ridiculed in mainstream media, which are represented in our model with assumption (i).

\begin{figure}[t]
\centering \includegraphics[scale=0.99]{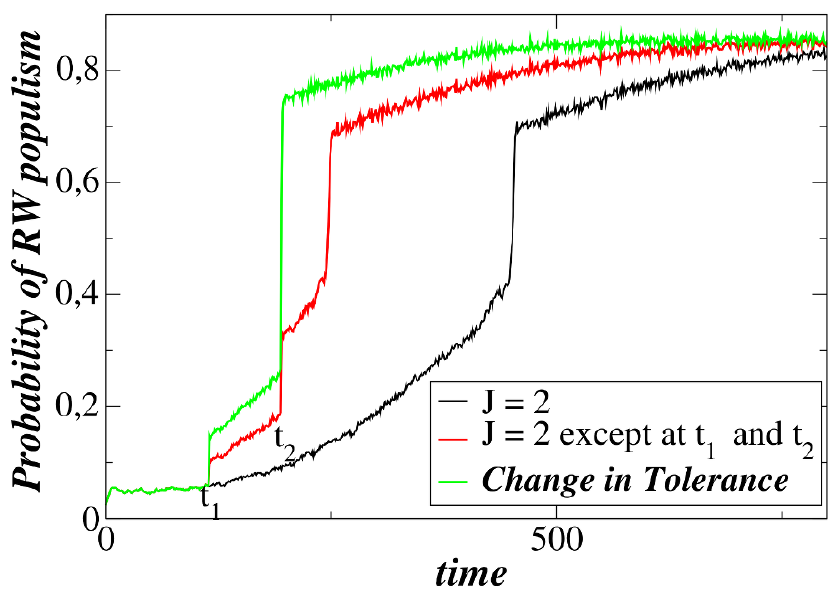}
\caption{\textbf{Non-linearity, a tipping point, and the rise of right-wing populism.} Using a network of $N=5\,000$ agents, each with an average of 15 connections, we examine the effect of a constant inflow of outsiders at rate $J=2$ at each time step. In this setup, the total number of outsiders at any moment in time is $I(t)=\Sigma_i J_i=\langle J\rangle t$. As the fraction of outsiders, $f_I=I/(N+I)$, approaches the tolerance parameter, $f_I'=0.15$, the presence of a tipping point causes the fraction of RW populist supporters to start increasing non-linearly and eventually undergo a sudden jump (i.e., a discontinuous change) at about 37\,years (450 months) into the simulation (black curve). The sudden jump happens much earlier if the inflow of outsiders experiences shocks at times $t_1$ and $t_2$ at which $J=200$ outsiders enter the network. In particular, as the network approaches the tipping point, the effect of exactly the same shock becomes disproportionately higher (red curve). In this case, however, the tolerance parameter is still kept constant. Finally, we also examine the case in which shocks at times $t_1$ and $t_2$ affect the tolerance parameter, where responsiveness is controlled by parameter $\gamma=0.0001$. Here, the second shock at $t_2$ is sufficient to instantly tip the network into RW populism (green curve). Other parameters are $p=0.007$, $\tau=15$, $p'=0.5$, $p''=0.5$, and $\alpha=0.001$.}
\label{f4}
\end{figure}
 
Assumption (iii) concludes model construction, thus letting us to refocus on the simulation results and the implications thereof. In Fig.~\ref{f4}, starting with a network of 5\,000 agents, we numerically show how the fraction of RW populist supporters increases with a constant inflow of outsiders, here $J=2$ per month. This inflow corresponds annually to almost 0.5\% of the total population, which is slightly more than the threshold value implied by Fig.~\ref{f3}(b). Simulations with $J=2$ per month are supposed to emulate the fast limit represented in extension (B) above. After approximately 37 years of fast globalization, the network reaches a tipping point and undergoes an abrupt shift to a mode dominated by RW populism. Stating that RW populism dominates implies more than 50\% of RW populist supporters in the network (i.e., $P>0.5$, where $P$ denotes the fraction of RW populists). The 50\% threshold is due to democracy as the majority rule principle.

In Fig.~\ref{f4}, we furthermore report how the simulated network under assumptions (i)-(iii) responds to shocks (red curve). At this stage, extension (B) is not yet allowed to operate. The constant annual inflow of outsiders, $J=2$, is supplemented with two events at times $t_1$ and $t_2$, at which $J=200$. The network's state, characterized by the proportion of RW populists ($P$), exhibits a much stronger response at $t_2$ than $t_1$, although the shock inflow ($J=200$) is the same. The reason for this stronger response is that in $t_2$ the system is closer to the tipping point and consequently more unstable than in $t_1$. Because in reality the value of $J$ may be biased as a consequence of erroneous estimation or some other form of information misinterpretation, the described results suggest that approaching the tipping point is concurrent with strengthening nonlinear effects such that even a small shock may trigger the final push to a mode dominated by RW populism. The situation is even more explosive (in terms of $P$) if extension (B) is allowed to operate, i.e., if tolerance parameter $f_I'$ changes with $J$.

\begin{figure}[t]
\centering \includegraphics[scale=0.99]{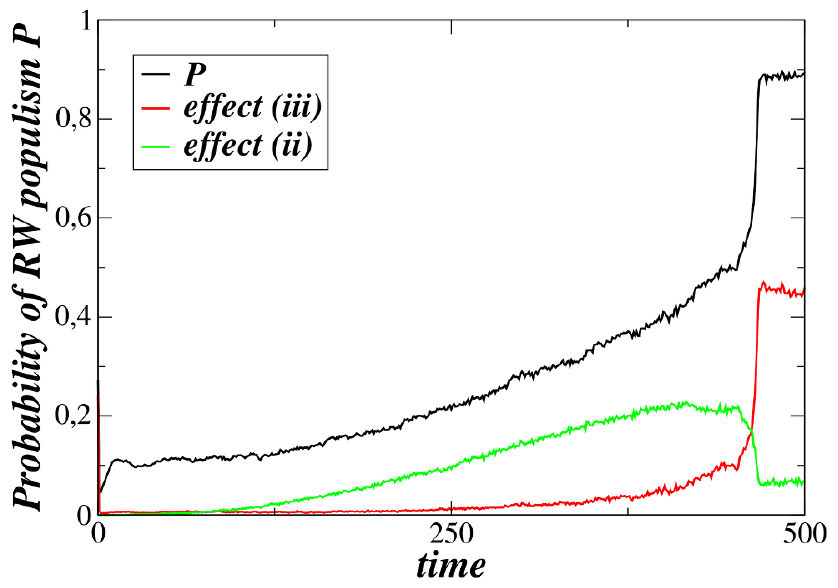} 
\caption{\textbf{Breakdown of the causes of right-wing populism}. Fig.~\ref{f4} shows that the probability of RW populism, $P$, suddenly increases as society approaches a tipping point, but remains silent on the underlying causes. Here we discern between the contributions of local outsider influence (assumption (ii)) and mutual insider contagion (assumption (iii)). Far from the tipping point, $P$ mainly responds to local outsider influence (ii). By contrast, as the network approaches its tipping point, mutual insider contagion (iii) takes over and accelerates the transition to RW populist dominance. Parameter values are $N=5\,000$ with an average degree of 15, $J=2$, $p=0.007$, $\tau=15$, $p'=0.7$, and $p''=0.8$.}
\label{f5}
\end{figure}

The third simulation in Fig.~\ref{f4} (green curve), illustrates the dynamics under assumptions (i)-(iii) with extension (B) operating. Accordingly, tolerance parameter $f_I$ changes with $J$ as prescribed by Eq.~(\ref{e3}). Due to decreasing tolerance, the second shock at $t_2$ is now sufficient to push the system beyond the tipping point and thus cause an even earlier dominance of RW populism than in previous two simulations.

From simulations in Fig.~\ref{f4} alone, it is unclear how much local influence of outsiders (assumption (ii)) contributes to the rise of RW populism relative to mutual insider contagion (assumption (iii)). We examine these contributions in Fig.~\ref{f5}. After the initial transients fade, it is the local influence of outsiders that drives the increase of RW populists in the network. The contribution of mutual insider contagion is relatively small until the system approaches a tipping point. Near the tipping point, contagion moves fast and overtakes the local outsider influence as the main contributor to the rise of RW populism. Thereafter, the RW populist movement can practically sustain itself without much support from the outside.

\begin{figure}[t]
\centering \includegraphics[scale=0.99]{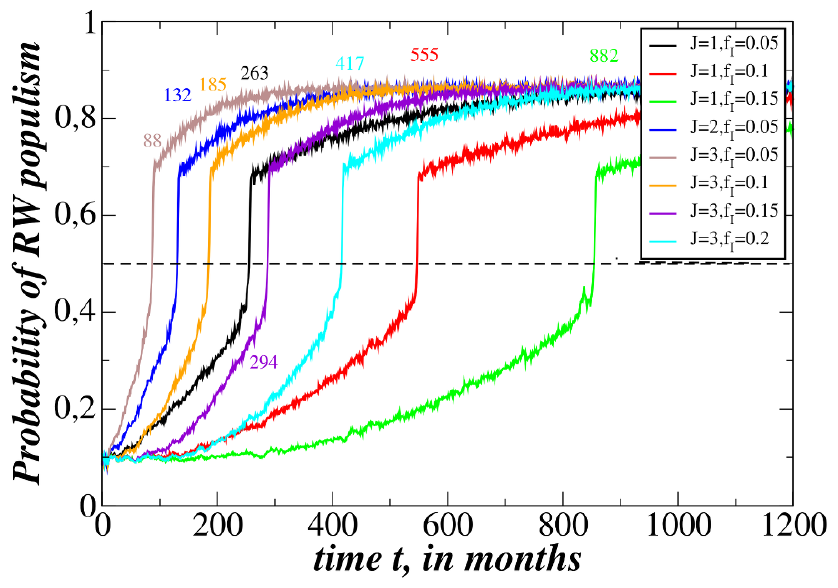} 
\caption{\textbf{Predicting the timing of RW populism}. We find that for a broad range of outsider inflows ($J$) and tolerance parameter values ($f_I'$), Eq.~(\ref{e5}) predicts the moment at which $P>0.5$ in a manner that agrees favorably with the simulation results. Except for $\gamma = 0$, other parameters are the same as in Fig.~\ref{f4}.}
\label{f6}
\end{figure}
   
In the regime of moderate immigration inflows ($J'<J<J''$), we can learn about the dynamics of our complex network using an analytical technique known in physics as the mean-field theory (MFT). As long as the number of agent connections does not deviate wildly from the network average, probability $P$ that a randomly chosen insider agent $i$ is an RW populist supporter due to any of the processes underlying assumptions (i)-(iii) is given by
\begin{eqnarray}
\nonumber 
P &=& p^* + p' p_1(k_i,m_i,f_I) + p'' p_1(K_i,M_i,P) \\
\nonumber 
  &-& p^* p' p_1(k_i,m_i,f_I) - p^* p'' p_1(K_i,M_i,P) \\
\nonumber 
  &-& p' p_1(k_i,m_i,f_I) p'' p_1(K_i,M_i,P),
\end{eqnarray}
where the last three terms avoid double counting in accordance with the probability theory formula $P(A\cup B\cup C)=P(A)+P(B)+P(C)-P(A) P(B)-P(A) P(C)-P(B) P(C)$ for three mutually independent events $A$, $B$, and $C$ that cannot occur simultanously. In the MFT approximation, we can drop index $i$ because no single agent is considerably different from the collective average. Previously we set $M=K/2$ for simplicity, but the peer pressure measured by the value of the proportionality factor between $M$ and $K$ may easily differ between countries or regions. Furthermore, parameters $p'$ and $p''$ are constants just in theory. The real social dynamics is such that these parameters may change in response to rumors, political manipulations, or outside shocks. If parameters $p'$ and $p''$ substantially increase, their effect is to increase the value of $P$ as well, thus further improving the prospects for the dominance of RW populism. In our framework, to reflect democracy as the majority rule principle, when $P$ approaches 0.5, the non-linear processes embedded in assumptions (ii) and (iii) lead to a sudden transition to RW populist mode.

\begin{figure}[t]
\centering \includegraphics[scale=0.99]{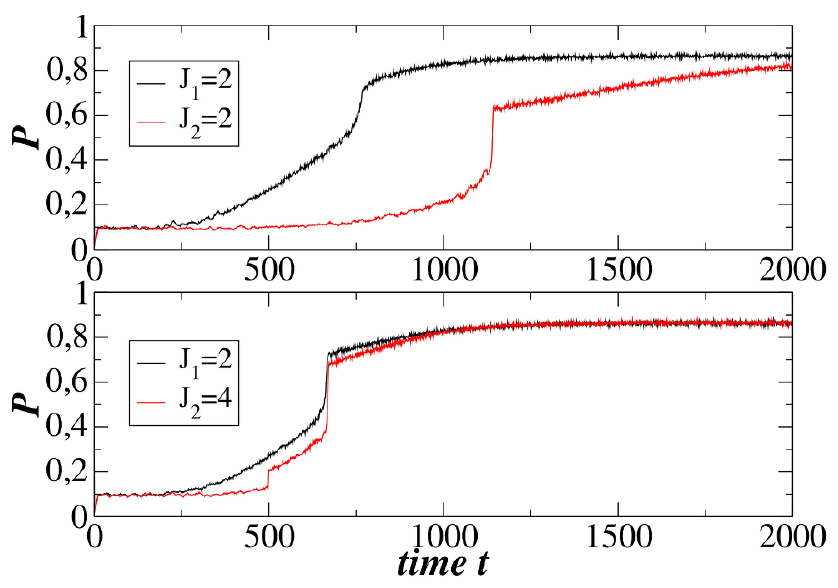} 
\caption{\textbf{Interconnected networks or why ``somebody else's problem'' easily turns into ``my problem''.} In (a) we show the case when there are no inter-links between networks. The tolerance parameters between the two networks differ, $f_{I,1} = 0.2$ and $f_{I,1} = 0.4$, while the inflows into both networks are the same, $J_1=J_2=2$. (b) More tolerant network is now exposed to a higher inflow, $J_2=4$, and a shock at $t_1=500$. The average number of connections for intra-connections (inter-connections) in both networks equals 15 (10). The other parameters are the same as in Fig.\ref{f4}.}
\label{f7}
\end{figure}

A theoretical model is more valuable if it possesses predictive power~\cite{Galam06}. We therefore demonstrate that a network of agents under assumptions (i)-(iii) leads to a simple formula for the timing at which RW populism starts to dominate. The formula in question is
\begin{equation} 
t_{th}=\frac{N f_I'}{J (1-f_I')}-\frac{I(0)}{J}.
\label{e5}
\end{equation}
We obtain this result in three steps. First, if the immigration inflow is constant, then the number of outsiders in the network after $t$ time steps is $I(0)+J t$. Second, the total population size thus equals $N+I(0)+J t$. Finally, Eq.~(\ref{e5}) follows if the current fraction of outsiders $(I(0)+J t)/(N+I(0)+J t)$ is equated with the critical parameter $f_I'$. In Fig.~\ref{f6}, we show for a number of immigration inflow--tolerance parameter pairs, $(J,f_I')$, that the simulated timing of the shift to RW populist mode (i.e., $P>0.5$) favorably fits theoretical predictions. In conjunction with the empirical data on tolerance towards immigrants in the EU countries, the formula in Eq.~(\ref{e5}) could be used to provide an estimate of when a given country might be approaching a possible tipping point at which RW populism becomes dominant.

Numerical simulations allow us not only to look at one network in isolation, but also to examine interdependence between two or even more networks. Especially interesting is the potential for a cascade effect whereby an RW populist movement in one network affects the rise of RW populism in another network. Such a cascade effect is interesting in a globalized world because globalization makes countries more similar to one another. This similarity should be even more present in supranational organizations like the EU wherein borders between nation states are all but erased. To examine how interdependence affects the rise of RW populism, we set up two random ER networks equivalent from an economic viewpoint (equal $p$ in the model), but different with respect to tolerance towards outsiders (different $f_I'$ in the model). An addition to the model is that, besides the usual intra-connections within one network, agents have inter-connections with their counterparts from the other network. Except this addition, the same model assumptions hold as before.

Numerical simulations reveal several interesting effects of interconnectedness. For easier comparison, we first run simulations in which inter-connections are erased (i.e., we just have two independent networks; Fig.~\ref{f7}(a)). As expected under the same inflows of outsiders ($J=2$), the network with a higher tolerance parameter ($f_{I,1}'=0.4$) approaches the tipping point much later than the network with a smaller tolerance parameter ($f_{I,2}'=0.2$). When the networks are interconnected and the more tolerant one experiences a higher inflow ($J_2=4$) and a shock ($J=500$) at time $t_1=500$, not only this network becomes more prone to RW populism, but it also pulls the other network, causing the transition to RW populism to occur sooner than it otherwise would (Fig.~\ref{f7}(b)). In essence, no one wants to be the first to cross the line, but in an interconnected world many may wait to be the second.
 
\section{Discussion and Conclusion}~Why some countries (e.g., the ex-socialist EU countries) strongly oppose receiving almost any immigrants, while others (e.g., the USA) have mostly welcomed immigration throughout history is an important topic in social sciences and, more recently, a major issue for the EU. Arguably one of the reasons why immigration has been welcomed in the USA is a lack of a single dominant ethnicity, leading to a clear distinction between the country's identity and the origin of its citizens. The second reason is that the USA has accepted people from all of the world, thus securing that there is no one large group of immigrants sharing religion, language, and/or ethnicity which could serve as a catalyst to mobilize this group and threaten the established social order. The opposite situation is in France where many immigrants share the same language and religion, both of which differ from the language and religion practiced by the French majority. A large homogeneous group of immigrants may instill fear among the current majority. Fear leads to a volatile situation, often exacerbated by the inflow rate of immigrants exceeding the speed of their integration. Such a situation is commonly resolved in one of the following two ways. Either there is an uprising as exemplified by the Visigoth immigrants and their ex-Roman commander Alaric who plundered Rome in 410, or the majority reacts to suppress the inflow which is nowadays accomplished through the support for populist right-wing parties.
 
The concept of globalization was conceived with cooperation between nations in mind, specifically to allow capital and labor to move freely across borders. In reality, however, globalization is affected by a multitude of non-economic factors such as ethnicity, culture, religion, and other human traits. For example, according to Eurobarometer 65 published in 2006, the main concerns of European citizens were related to unemployment (49\%), crime (24\%), the economic situation (23\%), immigration (14\%), and terrorism (10\%). However, a survey in the UK in 2006 lists race and immigration as a top issue mentioned by 38\% of the respondents, which may explain why populism showed up first in the UK. With so many factors at play, globalization is an enormously complex process and it should not come as a surprise if some of the non-economic factors fail to align with the purely economic ones. Such a misalignment is likely to lead to frustrations that feed populist movements around the world. By opposing collaboration in the spirit of globalization, populist movements may be creating a positive feedback mechanism whereby the general population becomes more ideologically rigid which, in turn, spurs populism even further~\cite{Jusup14}. A strengthened populist movement may even trigger tectonic shifts in world affairs as exemplified by BREXIT in the UK and the recent 2016 US presidential election.

What is sometimes forgotten is that in a globalized world problems rarely strike in one place only. Interdependence makes the developed countries more alike each other, thus synchronizing their social dynamics. This synchronization may help spill over political shifts in one country to other countries, eventually causing the rise of RW populism across large regions of the world. After BREXIT and the 2016 US presidential election, political elites should not expect to continue business as usual because being the first to undergo a major political shift (with almost certain unpleasant economic consequences) is difficult, but once someone has crossed that line in an interdependent world, a cascade (domino) effect may ensue---no one wants to be the first, but many wait to be the second.

A tipping point for the rise of RW populism may be easier to reach when voters face a binary choice. For example, such was the choice in the UK during the BREXIT referendum and, for all practical purposes, in the recent US presidential elections. In both cases the populist option secured a narrow victory. We also mentioned that in Austria, the RW populist party is currently receiving 34\% of votes in the polls, but its candidate in the presidential race is close to 51\%. A simple way to understand these percentages is to assume that political attitudes of voters are almost symmetrically distributed across the political spectrum from left to right. Consequently, if leftist voters comprise $\psi_L\%$ of the total population, then RW populist voters should maintain a similar presence, i.e., $\psi_R\%\approx\psi_L\%$. Facing only a binary choice, the remaining centrist voters are left without a clear representative of their views. These centrist voters are therefore likely to vote evenly between the options that are available. An implication is that around $\psi_L\%\approx\psi_R\%\approx33\%$, even a slight (statistically significant) imbalance in favor of $\psi_R\%$ over $\psi_L\%$ may tip a society towards RW populism. In Austria $\psi_R\%\approx34\%$ seems to be enough to promote the RW populist candidate into a front runner of the presidential race. In this context, polls from August 2016 indicate that in France the RW populist candidate for president has the support of about 28\% of voters. This percentage is very close to highlighted $\psi_R\%\approx33\%$. If the RW populist candidate managed to increase his support to $\psi_R\%\approx33\%$ before the first round of elections, he would be all but guaranteed to receive close to 50\% in the second round, illustrating the tipping point at work.

The final outcome in a ``battle'' between conflicting factors surrounding globalization will almost surely have tremendous economic implications. If a country approaches its tipping point, how the ensuing volatility will affect this country's credit rating in the long run? In the case of a domino effect, either in the entire EU or its large part, what will be the consequences for the Euro and the common banking system? Without a proper resolution of the migrant crisis, what will be the impact on systemic risk? In an attempt to shed light on some of the factors and underlying processes that affect the success of globalization, we offer an empirically motivated theoretical model of when to expect the rise of RW populism in response to unsustainable immigration inflows. 
 
It is an ongoing debate under what conditions globalization, democracy, and national sovereignty are able to coexist~\cite{Dani00, Dani16}. In our model, the coexistence of this trilemma set is predicated on such controlled globalization that a delicate balance is struck between immigration inflows and the ability of a society to integrate immigrants. This ability is arguably improved if immigrants mix more uniformly with the native population, which is a principle practiced in Singapore where immigrant hubs are discouraged and tenants in government-built housing (comprising 88\% of all housing) must
be of mixed ethnic origin~\cite{BP16}. Because tolerance towards immigrants is conditional, if immigration inflows overshadow integration rates, then democracy might push the society into RW populism as an intolerant mode of functioning. The rise of RW populism is possible because elections are stochastic in their very nature, implying that the society resembles a mathematical random walker. Left to its own devices, a random walker will eventually hit an absorbing barrier. This barrier, of course, is used here as a metaphor for the demise of globalization at the very hands of a progressive system (i.e., democracy) that made globalization possible in the first place.

\vspace{3mm}  
   
\noindent We are grateful to S. Galam, D. Kovac, and T. Lipic for helpful suggestions. B.~P. received support from the University of Rijeka. B.~P. and H.~E.~S. received support from the Defense Threat Reduction Agency (DTRA), the Office of Naval Research (ONR), and the National Science Foundation (NSF) Grant CMMI 1125290. M.~J. was supported by the Japan Science and Technology Agency (JST) Program to Disseminate Tenure Tracking System.

\end{document}